\documentclass[preprint,nofootinbib,showkeys,showpacs,tightenlines,fleqn]{revtex4}
\usepackage{amssymb} 
\usepackage{graphicx} 
\newcommand{\bee}{\begin{eqnarray}} 
\newcommand{\eee}{\end{eqnarray}} 
\newcommand{\nn}{\nonumber} 
\begin{document} 
%------------------------------------------------- 
\preprint{PNU-NTG-03/2006} 
\preprint{PNU-NURI-04/2006} 
\title{Twist-3 pion and kaon distribution amplitudes from the instanton 
vacuum with flavor SU(3) symmetry breaking} 
\author{Seung-il Nam} 
\email{sinam@pusan.ac.kr} 
\affiliation{Department of Physics and Nuclear Physics \& Radiation 
Technology Institute (NuRI), Pusan National University, Busan 609-735, 
Republic of Korea}  
\author{Hyun-Chul Kim} 
\email{hchkim@pusan.ac.kr} 
\affiliation{Department of Physics and Nuclear Physics \& Radiation 
Technology Institute (NuRI), Pusan National University, Busan 609-735, 
Republic of Korea}  
%------------------------------------------------- 
\begin{abstract} 
We investigate the twist-3 pion and kaon distribution amplitudes of the 
pseudoscalar ($\phi^{p}_{\pi,K}$) and pseudotensor ($\phi^{\sigma}_{\pi,K}$) 
types, based on the effective chiral action from the instanton vacuum. 
Flavor SU(3) symmetry breaking effects are explicitly taken into account. 
The Gegenbauer moments and the moments of the distribution amplitudes 
($\langle\xi^m\rangle$) are also computed. Our results are summarized as  
follows: $a^p_{2,\pi}\sim0.4$, $a^p_{1,K}\sim0.02$ and $a^p_{2,K}\sim0.14$, 
and $a^{\sigma}_{2,\pi}\sim0.02$ and $a^{\sigma}_{1,K}\sim 
a^{\sigma}_{2,K}\sim0$; 
$\langle\xi^2\rangle^p_{\pi}\sim\langle\xi^2\rangle^p_K\sim0.37$ and 
$\langle\xi\rangle^p_K\sim0$, and 
$\langle\xi^2\rangle^{\sigma}_{\pi}\sim\langle\xi^2\rangle^{\sigma}_K\sim0.20$  
and $\langle\xi\rangle^{\sigma}_K\sim0$. We compare our results with those  
from the QCD sum rules. We also discuss the relevant Wilson coefficients 
which were analyzed recently in chiral perturbation theory. 
\end{abstract} 
%------------------------------------------------- 
\maketitle 
%------------------------------------------------- 
\section{Introduction} 
%------------------------------------------------- 
The meson light-cone distribution amplitude (DA) plays an essential role in 
describing exclusive hadronic reactions~\cite{Efremov:1979qk, 
Lepage:1979zb,Lepage:1980fj,Chernyak:1983ej}. While the leading-twist DAs 
provide major contribution to those processes, higher-twist DAs are 
suppressed by the inverse of the momentum transfer $Q^{2}$, so that 
higher-twist DAs are less significant in studying high-energy exclusive 
processes. Because of this fact, not many investigations on higher-twist DAs 
were performed~\cite{Braun:1988qv,Ball:1998je, 
Huang:2004tp,Huang:2005av,Chen:2003fp,Chen:2005js,Ball:2006wn}, compared to 
the leading-twist ones. However, higher-twist DAs should be still important, 
since they give crucial information on the soft part of exclusive hadron 
reactions, in particular, in the smaller $Q^{2}$ region. Moreover, it 
describes the transverse motion of quarks inside the 
meson~\cite{Ball:1998je,Ball:2006wn}.   
 
In the present work, we aim at investigating the two-particle twist-3 
pion and kaon DAs within the framework of the nonlocal chiral quark 
model ($\chi$QM) from the instanton vacuum, taking into account SU(3) 
symmetry breaking effects. The instanton model of the QCD  
vacuum provides a good framework to study the structure of light 
mesons, since spontaneous chiral symmetry breaking is well realized in 
the instanton vacuum via the quark zero mode~\cite{ 
  Shuryak:1981ff,Diakonov:1983hh,Diakonov:1985eg}. This instanton  
vacuum model was later extended by introducing the current-quark 
masses~\cite{Musakhanov:1998wp,Musakhanov:2001pc,Musakhanov:vu}. It 
was assumed in the model that the large $N_c$ expansion is the 
reasonable one and the results were obtained in the leading order in 
this expansion. In the present approach, we employ the modified 
effective chiral action with flavor SU(3) symmetry breaking effects derived 
from the instanton vacuum~\cite{Musakhanov:1998wp,Musakhanov:2001pc, 
  Musakhanov:vu}.   
 
The model has been applied to describe the leading-twist meson 
DAs~\cite{Petrov:1998kg,Dorokhov:2002iu,Praszalowicz:2001wy, 
Praszalowicz:2001pi,Nam:2006au}. One of the two-particle twist-3 pion 
DAs was already studied within the nonlocal $\chi$QM in the chiral 
limit~\cite{Praszalowicz:2001pi}. There are two independent 
particle-two twist-3 distribution amplitudes defined as:  
\begin{eqnarray} 
\phi^p_{\mathcal{M}}(u)&=&\frac{\sqrt{2}(P\cdot\hat{n}) 
(m_f+m_g)}{m^2_{\mathcal{M}}F_{\mathcal{M}}}\int^{ 
\infty}_{-\infty}\frac{d\tau}{\pi}\, e^{-i(2u-1)\tau P\cdot\hat{n}} 
\langle0|\bar{\psi}_f(\tau\hat{n})i\gamma_5\psi_g(-\tau\hat{n})| 
\mathcal{M}(P)\rangle,  \label{DAS} \\ 
\phi^{\sigma}_{\mathcal{M}}(u)&=&-\frac{6\sqrt{2}(m_f+m_g)}{ 
m^2_{\mathcal{M}}F_{\mathcal{M}}}\int^{\infty}_{-\infty} 
\frac{d\tau}{\pi} \int_0^udv\,e^{-i(2v-1)\tau P\cdot\hat{n}}  \cr 
&&\hspace{3cm}\times\langle0|\bar{\psi}_f( \tau\hat{n})i(\rlap{/}{P} 
\rlap{/}{\hat{n}}-P\cdot\hat{n}) \gamma_5\psi_g(-\tau\hat{n})| 
\mathcal{M}(P)\rangle, 
\label{DAT}   
\end{eqnarray} 
where the subscript $\mathcal{M}$ denotes either the pion 
($\mathcal{M}=\pi$) or the kaon ($\mathcal{M}=K$) with on-mass-shell 
momentum $P$ in the light-cone frame. $u$ and $n^{\mu}$ stand for the 
longitudinal momentum fraction and the light-like vector satisfying 
$n^2=0$, respectively.  The spatial separation between the quarks 
inside the meson is represented by $\tau\cdot n $. In the present 
work, we drop the Wilson line by considering the light-cone gauge, 
$A\cdot n=0$.  $m_f$ and $m_g$ are the corresponding current 
quark masses with given flavors $f$ and $g$. As for the present case, we 
assign them as  
$(f,g,{\cal M})=(s,u,K^+)$ and $(d,u,{\pi}^+) $. $m_{\mathcal{M}}$ 
denotes the corresponding meson mass.  $F_{\mathcal{M}}$ is the 
corresponding meson decay constant which is the normalization constant  
for the DAs.  The DAs of 
Eqs.~(\ref{DAS}) and (\ref{DAT}) satisfy the following normalization 
conditions: 
\begin{eqnarray} 
\int_0^1du\,\phi^p_{\mathcal{M}}(u)=\int_0^1du\,\phi^{\sigma}_{ 
\mathcal{M}}(u)=1.  \label{normalization} 
\end{eqnarray} 
 
The present work is organized as follows: Section II is devoted to the 
general formalism. The numerical results and discussions are given in 
Section III. We summarize and draw conclusions in Section IV. 
%-------------------------- 
\section{General Formalism} 
%-------------------------- 
In order to evaluate the nonlocal hadronic matrix elements of 
Eqs.~(\ref{DAS}) and (\ref{DAT}), we start from the low-energy 
effective QCD partition function derived from the instanton vacuum in 
Euclidean space as 
follows~\cite{Diakonov:1985eg,Musakhanov:1998wp,Musakhanov:2001pc, 
Musakhanov:vu}: 
\begin{eqnarray} 
\mathcal{Z} &=&\int \mathcal{D}\psi \mathcal{D}\bar{\psi}\mathcal{D} 
\mathcal{M}^{a}\exp \int d^{4}x\Big[i{\psi}^{\dagger}_{f}(x)(i\rlap{/}{ 
\partial }+im_{f})\,\psi _{f}(x)  \nonumber \\ 
&&\hspace{4cm}-\int \frac{d^{4}k\,d^{4}p}{(2\pi )^{8}}e^{i(k-p)\cdot x}\sqrt{ 
M_{f}(k)M_{g}(p)}\,\bar{\psi}_{f}(k)U_{fg}^{\gamma _{5}}\,\psi 
_{g}(p)\Big],   
\label{effectiveaction}   
\end{eqnarray} 
where $\psi$ and ${\cal M}$ are the fields for the quark and pseudoscalar 
meson. $M_f(k)$ is the dynamically generated quark mass discussed later. The 
background pseudoscalar meson field $U^{\gamma _{5}}$ is given by   
\begin{equation} 
U^{\gamma _{5}}=U(x)\frac{1+\gamma _{5}}{2}+U^{\dagger }(x)\frac{1-\gamma 
_{5}}{2}=1+\frac{i}{F_{\mathcal{M}}}\gamma _{5}\mathcal{M}^{a} 
\lambda^{a}-\frac{1}{2F_{\mathcal{M}}^{2}}(\mathcal{M}^{a})^{2}\cdots .   
\end{equation} 
$\lambda ^{a}$ is the well-known SU(3) Gell-Mann matrices. Now we are in a 
position to construct the quark propagator in the instanton ensemble. There 
have been  
several approaches for this purpose~\cite{Musakhanov:1998wp, 
Musakhanov:2001pc,Musakhanov:vu}. In the present work, we make use of the 
prescription suggested by Pobylitsa~\cite{Pobylitsa:1989uq} and 
Musakhanov~\cite{Musakhanov:2001pc}. We attempt to discuss it  
briefly in order. First, Diakonov {\it et al.} 
introduced the quark propagator under the instanton  
effects by taking into account the following assumption~\cite{Diakonov:1995qy}: 
\begin{equation} 
\langle 
x|\frac{1}{i\rlap{/}{\partial}+\rlap{/}{A}_I+im_F}|x\rangle\simeq\langle  
x|\frac{1}{i\rlap{/}{\partial}+im_F}|x\rangle+\frac{\psi_I(x)\psi^{\dagger}_I(x)}{im_f}, 
\label{DPpro}   
\end{equation} 
where $A_I$ and $\psi(x)$ stand for the single instanton field and the quark 
zero mode, respectively. However, this assumption loses information on 
higher quark loops although it turns out that it works very well from 
phenomenological point  
of view. On the contrary, Pobylitsa expand the quark  
propagator $\langle x|(i\rlap{/}{\partial}+\rlap{/}{A}_I+im)^{-1}|x\rangle$ 
straightforwardly by virtue of the large $N_c$ 
limit~\cite{Pobylitsa:1989uq}. Thus, all non-planar  
diagrams are removed from the expansion. One then obtains an integral equation 
as follows:   
\begin{eqnarray} 
&&\langle x|\frac{1}{i\rlap{/}{\partial}+\rlap{/}{A}_I+im}|x\rangle\nn\\&& 
=\langle x|\frac{1}{i\rlap{/}{\partial}+im_f}|x\rangle+\frac{N}{2VN_c} 
{\rm tr}_{c}\left[\int d^4Z_I\left[\langle x|(i\rlap{/}{\partial} 
+\rlap{/}{A}_I+im_f)|x\rangle-\frac{1}{\rlap{/}{A}}\right]^{-1} 
+(I\to\bar{I})\right], 
\label{full}   
\end{eqnarray} 
where $Z_{I,\bar{I}}$ indicates the instanton coordinate for the instanton 
and antiinstanton. By solving Eq.~(\ref{full}) using an ansatz $\sigma(k)$ for 
the full propagator, $[i\partial+im_f+\sigma(i\partial)]^{-1}$, one can obtain 
the current-quark mass dpependent  
dynamical quark mass~\cite{Pobylitsa:1989uq,Musakhanov:2001pc}. Final 
expression for it can be written as follows: 
\begin{equation} 
M_f(k)=M_0F^2(k\bar{\rho}) 
\left[\sqrt{1+\frac{m^2_f}{d^2}}-\frac{m_f}{d}\right], 
\,\,\,\,{\rm where}\,\,\,\,  
d=\sqrt{\frac{0.08385}{2N_c}}\frac{8\pi\bar{\rho}}{R^2}\simeq0.198\,{\rm GeV}.   
\label{eq:mofk} 
\end{equation} 
We notice that, as a consequence, all the effects from the current-quark mass 
are included in the bracket of Eq.~(\ref{eq:mofk}).  Here, we employ the 
standard values for the instanton ensemble, $N/V=200^4$ MeV$^4$ and 
$\bar{\rho}\sim1/3$ fm $\simeq1/600$ MeV$^{-1}$ for the numerical 
calculations. From these values, we obtain $M_0=0.350$ GeV. Although the form 
factor $F(k)$ can be  
derived from the Fourier transform of the quark zero mode, we will use 
the following simple parameterization for it:   
\begin{equation} 
M_{f}(k)=M_{0}\left[ \frac{n\Lambda ^{2}} 
{(n\Lambda^{2}-k^{2}+i\epsilon )}\right] ^{2n} 
\left[ \sqrt{1+\frac{m_{f}^{2}}{d^{2}}}-\frac{m_{f}}{d}\right].  
\label{dqm}     
\end{equation} 
Note that we are now working in Minkowski space, in which the DA are well 
defined, rather than in Euclidean one. In the present work as done in our 
previous work~\cite{Nam:2006au}, we assume that the 
partition function of Eq.~(\ref{effectiveaction}) can be analytically 
continued to Minkowski space. Although we do not prrovide a firm theoretical 
proof for this assumpmtion, it turns out that it works qualitatively well from 
the  
phenomenological point of 
view~\cite{Dorokhov:2002iu,Praszalowicz:2001wy,Praszalowicz:2001pi}. $\Lambda 
$ is the cutoff   
mass which can be taken as the scale parameter of the present work. We 
use $\Lambda \simeq 1.0$ GeV for the numerical calculation as 
in~\cite{Nam:2006au}. In order to check the effect of the    
parameterization of the form factor, we take three different values 
of the power in Eq.~(\ref{dqm}), namely, 
$n=1$, $2$ and $3$. With the scheme discussed above, we can write the 
expressions for the  
two-particle twist-3 pion and kaon DAs, $\phi _{\mathcal{M}}^{p}$ and 
$\phi _{\mathcal{M}}^{\sigma }$:   
\begin{eqnarray} 
\phi_{\mathcal{M}}^{p}(u)&=&-\frac{iN_cP_+(m_{f}+m_{g})} 
{m_{\mathcal{M}}^{2}F_{\mathcal{M}}^{2}}  
\int \frac{dk_{+}dk_{-}dk_{T}^{2}}{(2\pi 
)^{4}}\delta \lbrack uP_{+}-k_{+}\rbrack  \nonumber \\ 
&{}&\hspace{3cm}\times \mathrm{tr}_{d}\left[\frac{\sqrt{M_{f}(k)}}{ 
D_{f}}\gamma _{5}\frac{\sqrt{M_{g}(k-P)}}{D_{g}}\gamma _{5}\right] ,   
\label{DAS1} 
\\ 
\phi _{\mathcal{M}}^{\sigma }(u)&=&\frac{6iN_c(m_{f}+m_{g})}{m_{ 
\mathcal{M}}^{2}F_{\mathcal{M}}^{2}}\int_{0}^{u}dv\int \frac{ 
dk_{-}dk_{+}d^{2}k_{T}}{(2\pi )^{4}}\delta \lbrack vP_{+}-k_{+} 
\rbrack  \cr 
&{}&\hspace{3cm}\times \mathrm{tr}_{d}\left[\frac{\sqrt{M_{f}(k)}}{D_{f}}( 
\rlap{/}{P}\rlap{/}{\hat{n}}-P_+)\gamma _{5}\frac{\sqrt{M_{g}(k-P) 
}}{D_{g}}\gamma _{5}\right],  \label{DAT1} 
\end{eqnarray} 
where $P_+=P\cdot\hat{n}$. $D_f$ denotes the inverse quark propagator, 
$\rlap{/}{k}+m_f+M_f(k)$. The trace $\mathrm{tr}_d$ denotes the trace over  
Dirac space. The explicit  
evaluation of the DAs in  
Eqs.~(\ref{DAS1}) and (\ref{DAT1}) is  
given in Appendix.  We set $m_{\pi}=140$ MeV and $m_K=495$ MeV for 
numerical input.  We assume isospin symmetry with $m_u = 
m_d=5\,\,\mathrm{MeV}$.  We choose 
$m_s=150\,\,\mathrm{MeV}$ for the strange current quark mass.  
%------------------------------------------------- 
\section{Numerical results} 
%------------------------------------------------- 
In the present Section, we provide the numerical results of the twist-3 
pion and kaon DAs.  We fix the parameters, using the normalization 
conditions in Eq.~(\ref{normalization}). The method is given in Refs.~\cite 
{Nam:2006au} in detail.   
 
\begin{figure}[t] 
\begin{tabular}{cc} 
\includegraphics[width=7cm]{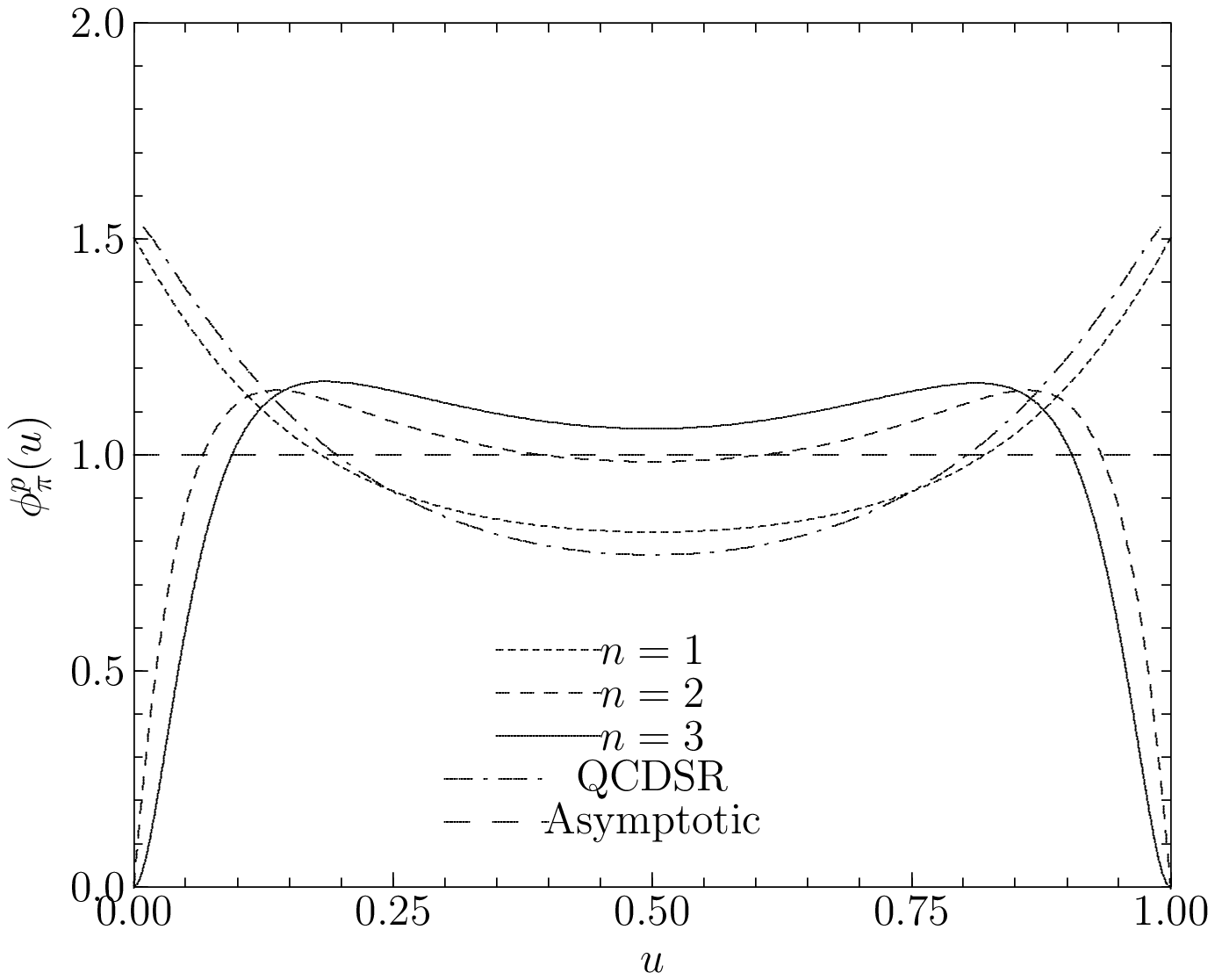}  
\includegraphics[width=7cm]{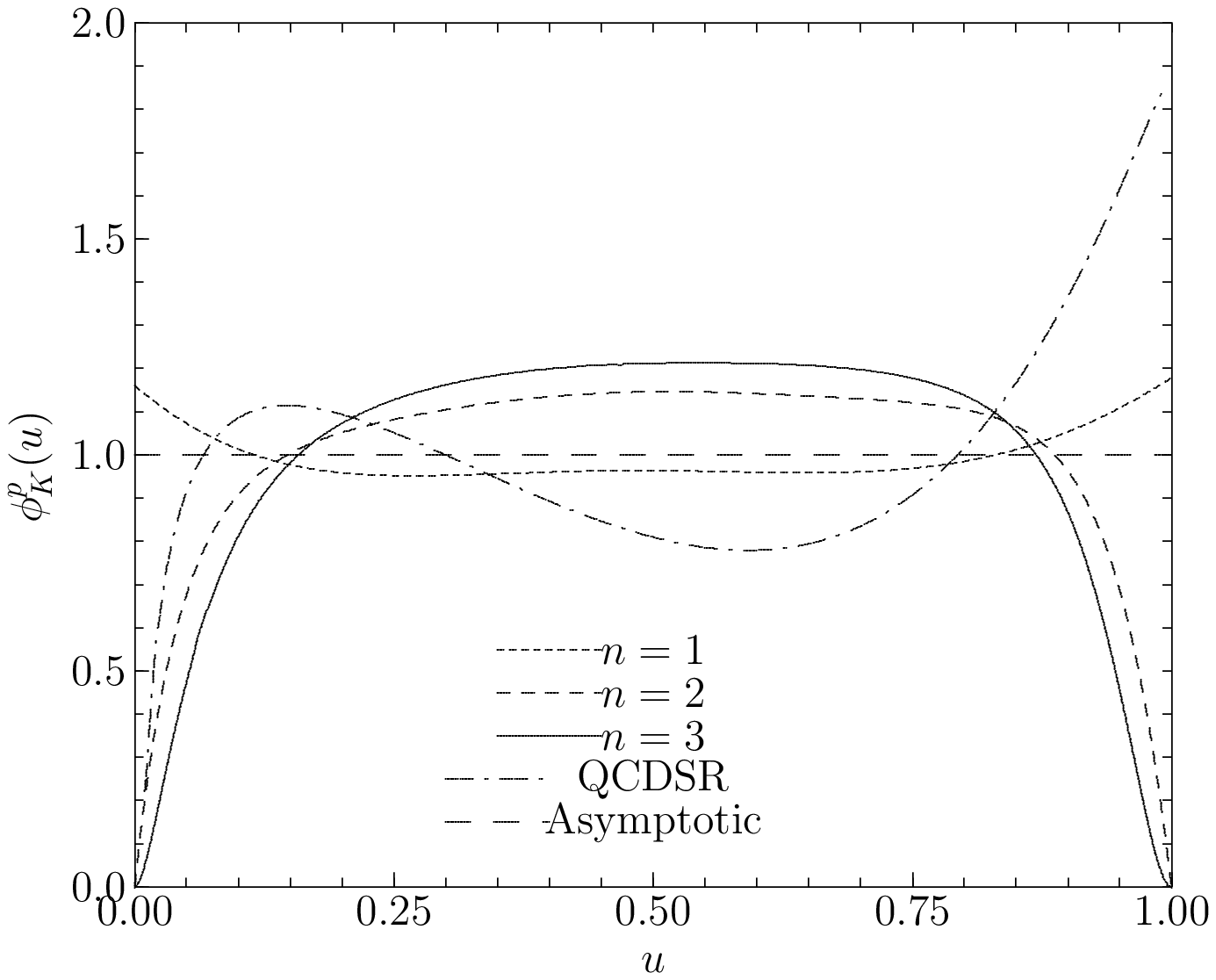} & 
\end{tabular} 
\caption{The results of the pesudoscalar twist-3 pion distribution 
amplitudes $\phi _{\pi }^{p}$ (left) and kaon 
ones $\phi _{K}^{p}$ (right). The  
dot-dashed curves indicate the results of the QCDSR in 
Ref.~\cite{Ball:2006wn}. The asymptotic one $\phi 
_{\mathrm{Asym.}}^{p}=1$ is drawn in the long-dashed line.  }     
\label{fig1} 
\end{figure}

In Fig.~\ref{fig1}, we depict the pseudoscalar-type DAs ($\phi 
_{\mathcal{M}}^{p}$) for the pion (left) and kaon (right). For 
comparison, we also show the asymptotic DA ($\phi_{\mathrm{Asym.}}^{p} 
(u)=1$) and those of Ref.~\cite{Ball:2006wn} in each panel.  Note that   
the results of Ref.~\cite{Ball:2006wn} using the QCD sum rules (QCDSR)  
were derived at the renormalization scale $\mu =1$ GeV, which is 
rather compatible to our case.  As expected from isospin symmetry,  
the pion DAs are all symmetric as shown in the left panel of 
Fig.~\ref{fig1}.  Examining the dependence of the DAs on the power $n$ 
in Eq.~(\ref{dqm}), we find a very interesting behavior of the DAs 
with power $n$.  With $n=1$ used, the pion DA does not vanish at the 
end points of $u$ and turns out to be very similar to that of 
Ref.~\cite{Ball:2006wn}, while the pion DAs are suppressed in the 
vicinity of the end points and eventually vanish at the end points in 
the case of $n=2$ and $3$.  In this case, the present results look 
similar to those of Ref.~\cite{Huang:2004tp}.  The reason lies in the 
fact that for the parameterization of the momentum-dependent quark mass 
with $n=1$ there exist nonvanishing terms at the end points after 
integration over $k_-$ variables.   
 
In the right panel, we depict the kaon DAs, $\phi_{K}^{p} (u)$.  
We see that the kaon DAs are almost symmetric in spite of the mass difference 
between the light and strangeness quarks. Note that this behavior was already 
seen even for the leading twist kaon DA in our previous 
work~\cite{Nam:2006au}.  It turns out that the present results of the 
$\phi_{K}^{p}$ are very   
different in shape, compared to that of Ref.~\cite{Ball:2006wn}. We conclude 
that, in our model framework, the pion and kaon DAs show negligible difference. 
\begin{figure}[t] 
\begin{tabular}{cc} 
\includegraphics[width=7cm]{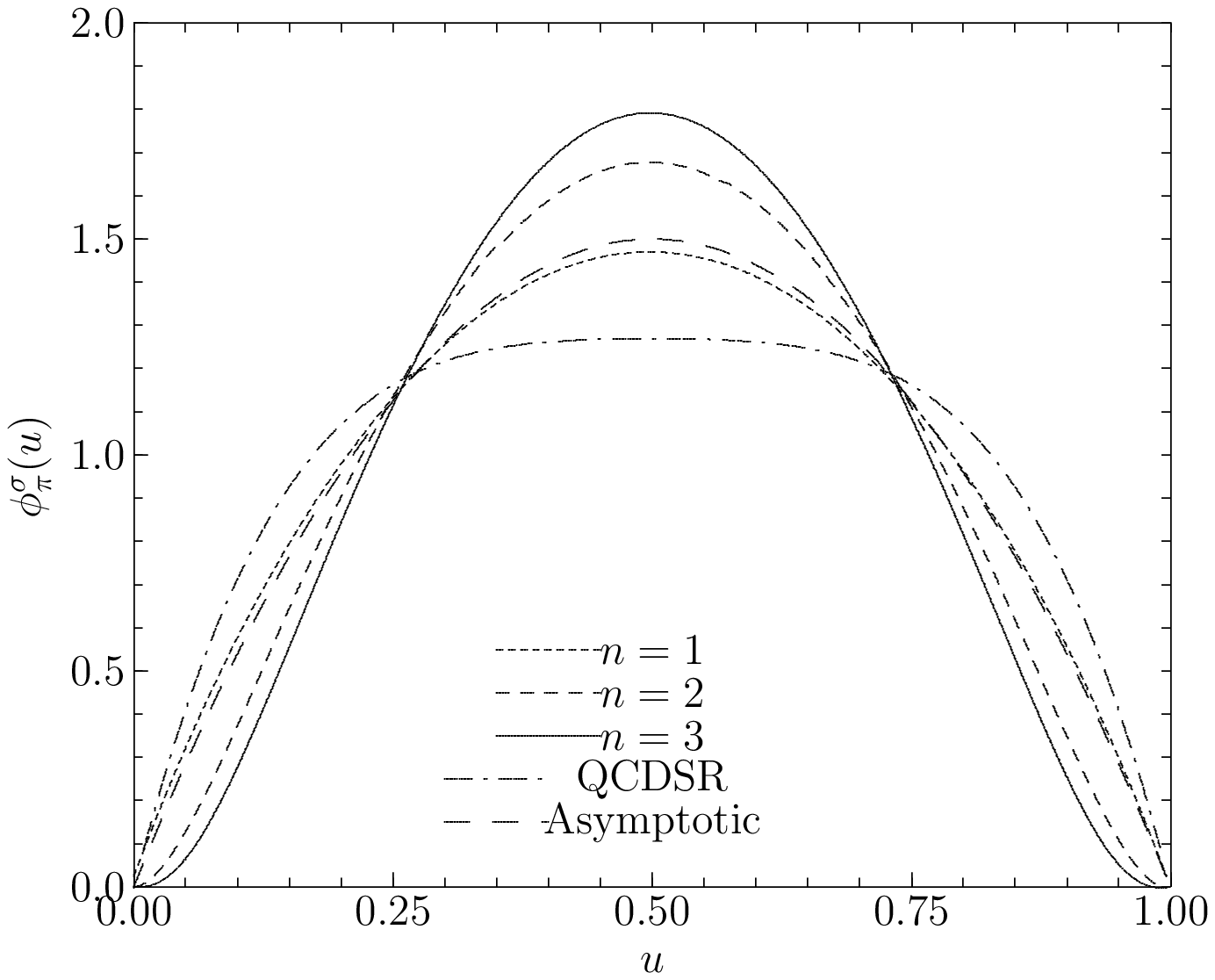}  
\includegraphics[width=7cm]{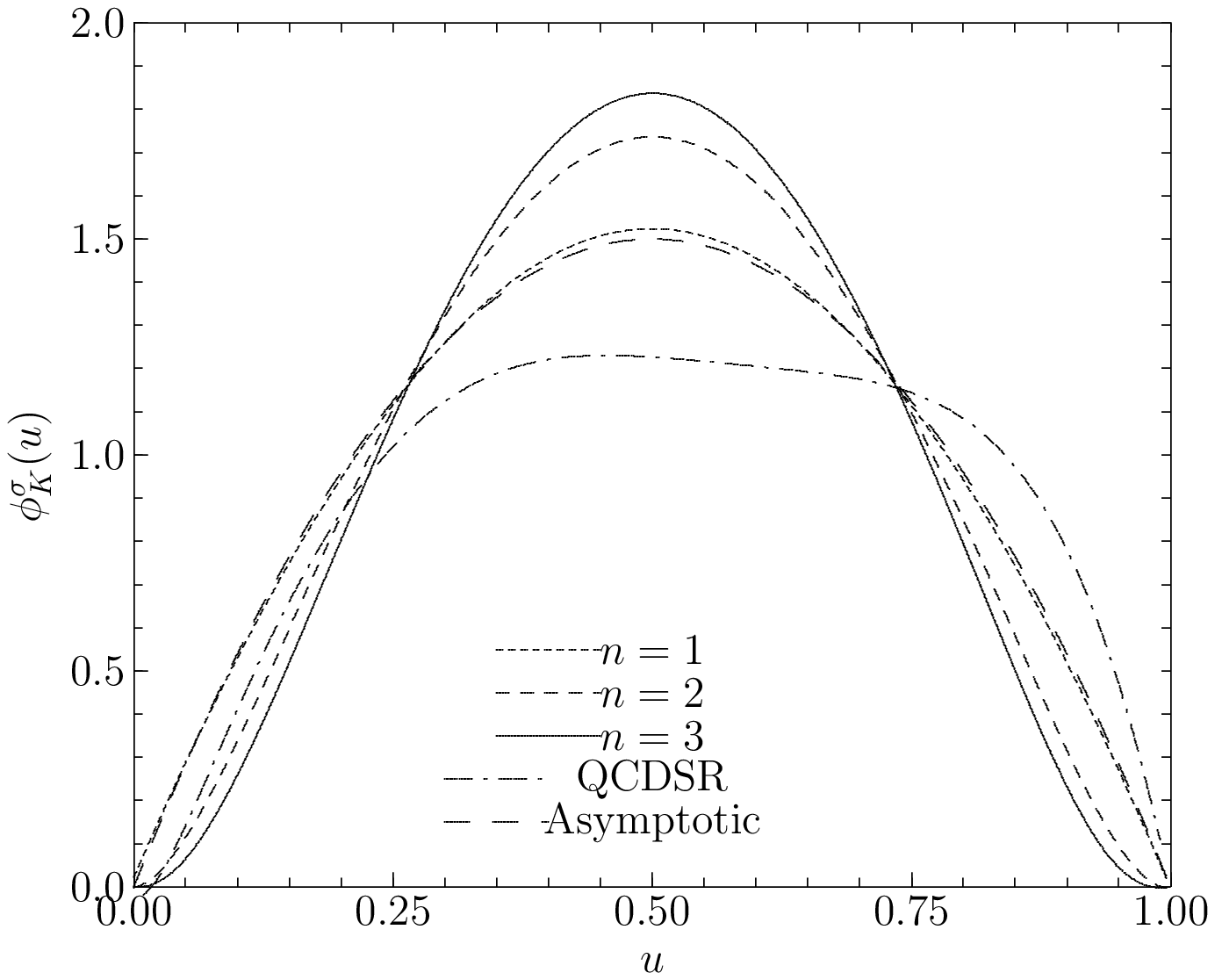}  
\end{tabular} 
\caption{The results of the pesudotensor twist-3 pion distribution 
amplitudes $\phi^{\sigma}_{\pi}$ (left) and kaon ones $\phi _{K}^{\sigma}$ 
(right). The dot-dashed curves indicate 
the results of the QCDSR  
in Ref.~\cite{Ball:2006wn}. The asymptotic one 
$\phi_{\mathrm{Asym.}}^{\sigma}=6u(1-u)$ is drawn in the 
dashed line. } 
\label{fig2} 
\end{figure} 
 
In Fig.~\ref{fig2}, we draw the present results of the pseudotensor 
twist-3 pion and kaon DAs ($\phi^{\sigma}_{\mathcal{M}}$) whose asymptotic 
form is $\phi_{\mathrm{Asym.}}^{\sigma}=6u(1-u)$. When $n=1$ is used, the 
$\phi^{\sigma}_{\pi}$ turns out to be  
almost the same as the asymptotic one.  However, as $n$ increases, the 
$\phi^{\sigma}_{\pi}$ are suppressed rather strongly near the end 
points, whereas they get narrower and larger in the 
neighborhood of the center ($u=0.5$).  It can be easily understood from 
the normalization condition for $\phi^{\sigma}_{\pi}$ in 
Eq.~(\ref{normalization}).  Compared to the result of 
Ref.~\cite{Ball:2006wn} again, the behavior of the present 
$\phi^{\sigma}_{\pi}$ is of great difference from it.  While the 
present $\phi^{\sigma}_{\pi}$ are humped near the center, that of 
Ref.~\cite{Ball:2006wn} is rather flat in that region.  The pesudotensor 
twist-3 kaon DAs $\phi^{\sigma}_{K}$ look similar  
to the pion ones, as indicated in the right panel of Fig.~\ref{fig2}, 
though they are negligibly asymmetric to the right. However, 
result of Ref.~\cite{Ball:2006wn} is shifted to the right and slightly 
asymmetric.   
     
In order to analyze the $\phi _{\mathcal{M}}^{p}$ and 
$\phi^{\sigma}_{\mathcal{M}}$ in detail, we compute their Gegenbauer 
moments.  By doing that, we can immediately check how much the 
DAs are deviated from the asymptotic ones. In addition, they give information 
on the strength of flavor SU(3) symmetry breaking effects for the kaon 
DAs. The twist-3 DAs can be  
expanded in terms of the Gegenbauer polynomials:   
\begin{eqnarray} 
  \label{eq:gegenp} 
\phi^p_{\cal M}(u)&=&\sum_{m=0}^{\infty}a^p_{m,\cal 
  M}C^{1/2}_m(\xi),\\  
  \label{eq:gegens} 
\phi^{\sigma}_{\cal M}(u)&=&6u(1-u)\sum_{m=0}^{\infty}a^p_{m,\cal  
  M}C^{3/2}_m(\xi),    
\end{eqnarray} 
where $\xi=2u-1$.  The orthogonal condition for the Gegenbauer 
polynomials, the Gegenbauer moments  
for the $\phi _{\mathcal{M}}^{p}$ and $\phi^{\sigma}_{\mathcal{M}}$  
can be derived as follows: 
\begin{eqnarray} 
&&a^p_{m,\mathcal{M}}=(2m+1)\int^1_0 du\, 
C^{1/2}_m(\xi)\,\phi^p_{\mathcal{M}}(u), \\  
&&a^{\sigma}_{m,\mathcal{M}}=\frac{4m+6}{3m^2+9m+6}\int^1_0 du\, 
C^{3/2}_m(\xi)\, \phi^{\sigma}_{\mathcal{M}}(u),  \label{gege2} 
\end{eqnarray} 
where $m$ denotes an order of the Gegenbauer moment.  The results of the 
Gegenbauer moments are listed in Table~\ref{table2}.  First, we 
consider the pseudoscalar ones, $a^p_{\mathcal{M}}$.  As expected from 
isospin symmetry, it turns out that all odd Gegenbauer moments vanish 
in the case of the pion DAs.  Moreover, we see that the Gegenbauer 
moments are all negative except for the case of $n=1$.  The result of 
$a_{2,\pi}^p$ is very similar in general to that of 
Ref.~\cite{Ball:2006wn}.  However, the values of $a_{4,\pi}^p$ and 
$a_{6,\pi}^p$ are positive in the present work but they 
are negative in Ref.~\cite{Ball:2006wn}, though in Fig.~\ref{fig1} 
they are seemingly almost the same each other.   
 
As for the kaon DA, we find that the situation becomes much more 
interesting.  The first Gegenbauer moment $a^p_{1,K}$,   
which encodes the strength of flavor SU(3) symmetry breaking effects, turns 
out to be all positive in our calculation.  Its value is about $0.02$. 
The smallness of the Gegenbauer moments explains why the 
kaon DA becomes almostly symmetric and rather flat.  Note that the even  
Gegenbauer moments are much smaller than those of the pion DAs.     
The results of the present work for the kaon DA are much smaller that 
those of the QCDSR~~\cite{Ball:2006wn}.   
\begin{table}[b] 
\begin{tabular}{c|ccc|cccccc}\hline 
$n$& $a^p_{2,\pi}$ & $a^p_{4,\pi}$ & $a^p_{6,\pi}$ & $a^p_{1,K}$ & $ 
a^p_{2,K}$ & $a^p_{3,K}$ & $a^p_{4,K}$ & $a^p_{5,K}$ & $a^p_{6,K}$ \\ \hline 
1& 0.44296 & 0.15081 & 0.08421 & 0.02033 & 0.13715 & 0.00170 & 0.10748 
& $-$0.00842 & 0.05975 \\  
2& $-$0.18451 & $-$0.45274 & $-$0.26330 & 0.02492 & $-$0.43993 & $-$ 
0.02802 & $-$0.32704 & $-$0.01548 & $-$0.16701 \\  
3& $-$0.43067 & $-$0.55590 & $-$0.17839 & 0.02364 & $-$0.64675 & $-$ 
0.03674 & $-$0.37241 & $-$0.01999 & $-$0.09403 \\ \hline 
\cite{Ball:1998je} & 0.51578 & 0.25454 & 0.21624 & $\cdots$ & 0.26310 & 
$\cdots$ & $-$  
0.05216 & $\cdots$ & 0.14697 \\  
\cite{Ball:2006wn} & 0.43726 & $-$0.07150 & $-$0.19686 & 0.18372 & 0.27067 &  
0.39530 & $-$0.24693 & 0.05497 & $-$0.24361 \\ \hline\hline 
$n$& $a^{\sigma}_{2,\pi}$ & $a^{\sigma}_{4,\pi}$ & $a^{\sigma}_{6,\pi}$ 
& $a^{\sigma}_{1,K}$ & $a^{\sigma}_{2,K}$ & $a^{\sigma}_{3,k}$ & $ 
a^{\sigma}_{4,K}$ & $a^{\sigma}_{5,K}$ & $a^{\sigma}_{6,K}$ \\ \hline 
1& 0.01658 & 0.00401 & 0.00230 & $-$0.00359 & $-$0.00735 & $-$0.00367 
& 0.00368 & $-$0.00280 & 0.00188 \\  
2& $-$0.09841 & $-$0.01923 & $-$0.00368 & $-$0.00474 & $-$0.11797 & $-$ 
0.00298 & $-$0.01314 & $-$0.00068 & $-$0.00282 \\  
3& $-$0.14770 & $-$0.01596 & $-$0.00052 & $-$0.00466 & $-$0.16072 & $-$ 
0.00335 & $-$0.00949 & $-$0.00002 & 0.00002 \\ \hline 
\cite{Ball:1998je} & 0.09400 & $-$0.00057 & $-$0.00078 & $\cdots$ & 0.05203 & $\cdots$ &  
$-$0.00048 & $\cdots$ & $-$0.00066 \\  
\cite{Ball:2006wn} & 0.09787 & $-$0.00156 & $-$0.00114 & 0.08976 & 0.05383 &  
0.05111 & $-$0.01501 & 0.00734 & $-$0.00578 \\  
\hline 
\end{tabular} 
\caption{Gegenbauer moments for the twist-3 pion and kaon 
  distribution amplitudes, $a^{p,\sigma}_{m,\cal M}$.} 
\label{table2} 
\end{table} 
 
We now consider the Gegenbauer moments for the 
$\phi^{\sigma}_{\mathcal{M}}$. The overall tendency for the sign of 
the Gegenbauer moments is very similar to that for the 
$\phi^p_{\mathcal{M}}$.  We find that the present results of the  
$a^{\sigma}_{2\pi}$ are much smaller than those of  
Ref.~\cite{Ball:1998je,Ball:2006wn}.  Lager values of the  
$a^{\sigma}_{2,\pi}$ imply that the DA becomes flat, as shown in 
Fig.~\ref{fig2}. Since we obtain all negative  
$a^{\sigma}_{2,K}$, the kaon DAs have narrower shapes.  Note that all 
odd kaon Gegenbauer moments are very small 
($\sim10^{-2}$).  Once again, we see the reduced SU(3) symmetry 
breaking effects.    
 
It is also of great interest to study the moments of the DAs defined 
as follows:  
\begin{eqnarray} 
&&\langle\xi^m\rangle^p_{\mathcal{M}}=\int^1_0du\, 
\xi^m\phi^p_{\mathcal{M}}(u), \\ 
&&\langle\xi^m\rangle^{\sigma}_{\mathcal{M}}=\int^1_0du\, 
\xi^m\phi^{\sigma}_{\mathcal{M}}(u).   
\label{momentsPS} 
\end{eqnarray} 
The results are listed in Table~\ref{table5}, being compared with the 
QCDSR results from Refs.~\cite{Ball:2006wn,Huang:2004tp,Huang:2005av}. 
We observe that $\langle\xi^{2,4,6}\rangle^p_{\pi}$ are comparable 
with the QCDSR results~\cite{Ball:2006wn,Huang:2004tp}. We find that the 
values of $\langle\xi^{1,3,5}\rangle^p_K$ are samll in  
comparison with Refs.~\cite{Ball:2006wn,Huang:2004tp}, although the   
even moments are still similar.  The results of  
Ref.~\cite{Huang:2005av} are slightly larger than the present ones in 
general.   
 
As for the moments of the $\phi^{\sigma}_{\mathcal{M}}$, it turns 
out that the present results are comparable with those of 
Ref.~\cite{Ball:2006wn}, though the difference between them is getting 
larger for higher moments.  The situation is similar when we 
compare our results with those of Ref.~\cite{Huang:2005av}.  We note 
that the odd moments become all negative.  
\begin{table}[b] 
\begin{tabular}{c|ccc|ccccc}\hline 
$n$& $\langle\xi^2\rangle^p_{\pi}$ & $\langle\xi^4\rangle^p_{\pi}$ & $ 
\langle\xi^6\rangle^p_{\pi}$ & $\langle\xi\rangle^p_{K}$ & $ 
\langle\xi^2\rangle^p_{K}$ & $\langle\xi^3\rangle^p_{K}$ & $ 
\langle\xi^4\rangle^p_{K}$ & $\langle\xi^5\rangle^p_{K}$ \\ \hline 
1& 0.39240 & 0.25445 & 0.19072 & 0.00678 & 0.35162 & 0.00416 & 0.21840 
& 0.00292 \\  
2& 0.30873 & 0.16742 & 0.10820 & 0.00831 & 0.27468 & 0.00338 & 0.14142 
& 0.00160 \\  
3& 0.27591 & 0.13666 & 0.08164 & 0.00788 & 0.24710 & 0.00263 & 0.11663 
& 0.00081 \\ \hline 
\cite{Huang:2004tp} & $0.340\sim0.359$ & $0.164\sim0.211$ & $\cdots$ &  
$\cdots$ & $\cdots$ & $\cdots$ & $\cdots$ & $\cdots$  \\  
\cite{Huang:2005av} & $0.52\pm0.03$ & $0.44\pm0.01$ & $\cdots$ & $-0.10\pm0.03$ & $ 
0.43\pm0.04$ & $\cdots$ & $\cdots$ & $\cdots$  \\  
\cite{Ball:2006wn} & 0.38653 & 0.24510 & 0.17879 & 0.06124 & 0.36757 &  
0.05933 & 0.22355 & 0.05198 \\  
\hline\hline 
$n$& $\langle\xi^2\rangle^{\sigma}_{\pi}$ & $\langle\xi^4\rangle^{ 
\sigma}_{\pi}$ & $\langle\xi^6\rangle^{\sigma}_{\pi}$ & $\langle\xi\rangle^{ 
\sigma}_{K}$ & $\langle\xi^2\rangle^{\sigma}_{K}$ & $\langle\xi^3\rangle^{ 
\sigma}_{K}$ & $\langle\xi^4\rangle^{\sigma}_{K}$ & $\langle\xi^5\rangle^{ 
\sigma}_{K}$ \\ \hline 
1& 0.20568 & 0.08992 & $-$0.00186 & $-$0.00216 & 0.19748 & $-$0.00162 
& 0.08442 & $-$0.00131 \\  
2& 0.16626 & 0.06122 & $-$0.00145 & $-$0.00284 & 0.15955 & $-$0.00179 
& 0.05738 & $-$0.00123 \\  
3& 0.14936 & 0.05030 & $-$0.00117 & $-$0.00280 & 0.14490 & $-$0.00184 
& 0.04799 & $-$0.00125 \\ \hline 
\cite{Huang:2005av} & $0.34\pm0.03$ & $0.20\pm0.01$ & $\cdots$ & $-0.13\pm0.04$ & $ 
0.173\pm0.002$ & $\cdots$ & $\cdots$ & $\cdots$ \\  
\cite{Ball:2006wn} & 0.23250 & 0.10747 & 0.06240 & 0.06124 & 0.20218 &  
0.03282 & 0.08949 & 0.02208 \\ \hline 
\end{tabular} 
\caption{Moments for the twist-3 pion and kaon distribution 
amplitudes, 
$\langle\xi^m\rangle^{p,\sigma}_{\mathcal{M}}$.}  
\label{table5} 
\end{table} 
 
Finally, we want to discuss the chiral perturbation theory ($\chi$PT) 
Wilson coefficients of Refs.~\cite{Chen:2003fp,Chen:2005js}, based on 
the present results\footnote{The analysis of the $\chi$PT Wilson 
coefficients can be performed by using the derivative expansion within 
the present scheme.  The corresponding work will be found 
elsewhere.}.  
It is of great interest to study them within the present model, since 
the model is chiral symmetric by construction.  In 
Refs.~\cite{Chen:2003fp,Chen:2005js}, it was shown that the  
non-local quark bilinear operators for the twist-3 meson   
DAs and their moments are reconstructed in terms of the chiral field, 
$\Sigma=\exp(i\sqrt{2}{\cal M}^a\lambda^a/F_{\cal M})$, to the leading order 
(LO) and next-to-leading order (NLO) as follows:   
\begin{eqnarray} 
\langle\xi^m\rangle_{\mathcal{M}}=\langle0|c_m\mathcal{O}_{\mathrm{LO}} 
+b_m \mathcal{O}_{\mathrm{NLO}}|\mathcal{M}(P)\rangle,  \label{op} 
\end{eqnarray} 
where $c_m$ and $b_m$ are the $\chi$PT Wilson coefficients encoding 
physics in the scale of $\Lambda_{\mathrm{\chi PT}}$.  From 
Eq.~(\ref{op}), one can derive the equations concerning the even and 
odd moments:   
\begin{eqnarray} 
&&\langle\xi^{2m+1}\rangle^{S,T}_{\pi}=0,  \nonumber \\ 
&&\langle\xi^{2m+1}\rangle^{S,T}_{K}=(m_s-\bar{m})b_{2m+1,2},  \nonumber \\ 
&&\langle\xi^{2m}\rangle^{S,T}_{\pi}=\langle\xi^{2m}\rangle^{S,T}_{m_q=0}+2 
\bar{m}\alpha_{2m}+(2\bar{m}+m_s)\beta_{2m},  \nonumber \\ 
&&\langle\xi^{2m}\rangle^{S,T}_{K}=\langle\xi^{2m}\rangle^{S,T}_{m_q=0}+( 
\bar{m}+m_s)\alpha_{2m}+(2\bar{m}+m_s)\beta_{2m}\,  \label{chpt} 
\end{eqnarray} 
where $\beta_{2m}$ and $\alpha_{2m}$ are the collective parameters of 
$c_m$ and $b_m$.   Note that these equations are model-independent, 
since they are only based on chiral symmetry of the 
operators~\cite{Chen:2003fp, Chen:2005js}.  The first equation in 
Eq.~(\ref{chpt}) indicates that the odd moments are all zero 
for the pion DAs, which is true because of isospin symmetry.  The 
second equation represents flavor SU(3) symmetry breaking  
effects in terms of the mass difference between the light and strange 
current quarks.  $\bar{m}$ is defined as $(m_u+m_d)/2=5$ MeV.  As for 
the even moments, the third and fourth equations are given.  They can 
be simplified due to the fact that $\phi^p_{\mathcal{M}}$ and 
$\phi^{\sigma}_{\mathcal{M}}$ are all zero in the chiral limit as 
easily verified in Eqs.~(\ref{DAS1}) and (\ref{DAT1}).  
\begin{eqnarray} 
&&\alpha_{2m}=\frac{\langle\xi^{2m}\rangle_{\pi}-\langle\xi^{2m}\rangle_{K}}{ 
\bar{m}+m_s}, \label{alpha} 
\\ 
&&\beta_{2m}=\frac{(\bar{m}+m_s)\langle\xi^{2m}\rangle_{\pi}-2\bar{m} 
\langle\xi^{2m}\rangle_{K}}{m^2_s+\bar{m}m_s-2\bar{m}^2}. 
\end{eqnarray} 
It is straightforward to obtain the values of $b_{2m+1}$ by using the 
results listed in Table~\ref{table5}.  The corresponding results are 
listed in Table~\ref{table6}.  
\begin{table}[t] 
\begin{tabular}{c|ccc|ccc}\hline 
$n$& $b^p_1$ & $b^p_3$ & $b^p_5$ & $b^{\sigma}_1$ & $b^{\sigma}_3$ & $ 
b^{\sigma}_5$ \\ \hline 
1& 0.04675 & 0.02872 & 0.02011 & -0.01487 & -0.01119 & -0.00900 \\  
2& 0.05730 & 0.02333 & 0.01105 & -0.01961 & -0.01231 & -0.00849 \\  
3& 0.05433 & 0.01812 & 0.00561 & -0.01929 & -0.01266 & -0.00860 \\  
\hline\hline 
$n$& $\alpha^p_2$ & $\alpha^p_4$ & $\alpha^p_6$ & $\alpha^{\sigma}_2$ &  
$\alpha^{\sigma}_4$ & $\alpha^{\sigma}_6$ \\ \hline 
1& 0.26310 & 0.23258 & 0.19845 & 0.05290 & 0.03548 & -0.31503 \\  
2& 0.21968 & 0.16774 & 0.12555 & 0.04329 & 0.02477 & -0.18723 \\  
3& 0.18587 & 0.12923 & 0.08890 & 0.02877 & 0.01490 & -0.14587 \\  
\hline 
\end{tabular} 
\caption{$\chi$PT Wilson coefficients $b_{2m+1}$ and 
$\alpha_{2m}$ defined in Eqs.~(\ref{chpt}) and (\ref{alpha}).} 
\label{table6} 
\end{table} 
Since the $b_{2m+1}$ are the expansion coefficients for the operators 
of NLO~\cite{Chen:2003fp,Chen:2005js}, we find that the effects from 
the NLO are getting weaker as $m$ increases. In addition to $b_{2m+1}$,  
we also list the calculated values of $\alpha_{2m}$ in 
Table~\ref{table6}.  We verify that the values of $\beta_{2m}$ are 
rather tiny ($10^{-3}$).  Note that $\beta_{2m}$ are the coefficients 
for the NLO.  It turns out that the values of the $\alpha_{2m}$ for 
the $\phi^p_{\mathcal{M}}$ are rather sizeable, while they are less 
than $10^{-2}$ for the $\phi^{\sigma}_{\mathcal{M}}$.  
%------------------------------------------------- 
\section{Summary and Conclusion} 
%------------------------------------------------- 
We investigated the two types of two-particle twist-3 pion and kaon 
distribution amplitudes, based on the nonlocal chiral quark model from  
the instanton vacuum.  We considered explicitly flavor SU(3) symmetry 
breaking effects, employing the modified effective chiral 
action~\cite{Musakhanov:2001pc,Musakhanov:vu}. The 
current-quark mass dependence was also taken into account for the dynamical 
quark mass $M_f(k)$ by solving the integral equation for the quark propagator 
under the instanton effects. We employed simple-pole  
type parameterization for the form factor in $M_f(k)$.  
 
Numerical results of $\phi^p_{\mathcal{M}}$ and 
$\phi^{\sigma}_{\mathcal{M}}$ were given for various cases.  We 
obtained the symmetric pion distribution amplitudes due to  
isospin symmetry.  On the contrary, the kaon distribution amplitudes 
turned out to be asymmetric because of SU(3) symmetry breaking effects. 
However, calculated curves for the kaon DAs were almost symmetric in spite of 
the quark mass differences.    
 
When power $n=1$ for the dynamical quark mass was used, we found that 
the results of the $\phi^{p}_{\mathcal{M}}$ seem to be very similar to 
those of Ref.~\cite{Ball:2006wn} with the same end-point behavior. 
However, as $n$ increased, the results look similar to those of 
Ref.~\cite{Huang:2004tp}.  As for the $\phi^{\sigma}_{\mathcal{M}}$, the 
results were very similar to that of the asymptotic one, with $n=1$ 
used.  As the power $n$ increases, the kaon distribution amplitudes 
were suppressed at the end points and were getting narrower.   
We also investigated the Gegenbauer moments and the moments of the 
distribution amplitudes. 
 
Finally, we estimated the $\chi$PT Wilson coefficients for the  
twist-3 pseudoscalar meson distribution amplitudes, using the 
present results.  These estimations may be useful in analyzing the 
higher-twist meson distribution amplitudes in terms of the chiral 
field operators.  A more systematic analysis can be carried out for 
the $\chi$PT Wilson coefficients, the derivative expansion being 
employed.  The corresponding analysis will appear soon.  
 
The investigation on the three-particle twist-3 and twist-4 
distribution amplitudes with gluon operators is under way.   
%------------------------------------------------- 
\section*{Acknowledgments} 
%------------------------------------------------- 
The present work has been supported by the two-year project of Pusan 
National University.  The work of S.i.N. is supported by the Brain 
Korea 21 (BK21) project in Center of Excellency for Developing Physics 
Researchers of Pusan National University, Korea. The authors would 
like to thank M.~M.~Musakhanov for his comments on this 
work. S.i.N. is grateful to Y.~Kwon for fruitful discussions.   
%------------------------------------------------- 
\section*{Appendix} 
%------------------------------------------------- 
\subsection{Twist-3 pseudoscalar type pseudoscalar meson DA} 
%------------------------------------------------- 
The twist-3 pseudoscalar type DA can be written by using 
Eq.~(\ref{DAS}) as follows:  
\begin{eqnarray} 
\phi^p_{\cal M}(u)&=&\frac{(P\cdot\hat{n})(m_f+m_g)}{m^2_{\cal M}f_{\cal 
    M}}\int^{\infty}_{-\infty}\frac{d\tau}{\pi}\,e^{-i(2u-1)\tau 
  P\cdot\hat{n}}\langle0|\bar{\psi}_f(\tau\hat{n})i\gamma_5\psi_g( 
-\tau\hat{n})|{\cal M}(P)\rangle\cr 
&&\hspace{-2cm} = \frac{\sqrt{2}(P\cdot\hat{n}) 
(m_f+m_g)}{m^2_{\cal M}F_{\cal 
  M}}\int^{\infty}_{-\infty}\frac{d\tau}{\pi} 
\,e^{-i(2u-1)\tau P 
\cdot\hat{n}}\langle0|\bar{\psi}_f(\tau\hat{n})i\gamma_5 
\psi_g(-\tau\hat{n})|{\cal M}(P)\rangle, 
\label{aps1}   
\end{eqnarray} 
where $F_{\cal M}=f_{\cal M}/\sqrt{2}$ stands for the empirical value of the 
pseudoscalar meson decay constant. We set $m_{\cal M}$ to be the mass of the 
pseudoscalar meson. After integrating over $\tau$ and $k_+$, we arrive at 
Eq.~(\ref{DAS1}). The trace shown in Eq.~(\ref{DAS1}) can be evaluated in the 
present framework as follows:   
\begin{eqnarray} 
&&{\rm tr}_{d}\left[\frac{\sqrt{M_f(k)}}{D(k)}\gamma_5 
\frac{\sqrt{M_g(k-P)}}{D(k-P)}\gamma_5\right]=4M_0 
\sqrt{f(m_f)f(m_g)}\left[\frac{n\Lambda^2}{n\Lambda^2-k^2} 
\right]^{n}\left[\frac{n\Lambda^2}{n\Lambda^2-(k-P)^2}\right]^{n} 
\nn\\&\times&\left\{k^2-k\cdot P-\left[m_f+M_0f(m_f)\left[ 
\frac{n\Lambda^2}{n\Lambda^2-k^2}\right]^{2n}\right]\left[ 
m_g+M_0f(m_g)\left[\frac{n\Lambda^2}{n\Lambda^2-(k-P)^2} 
\right]^{2n}\right]\right\}\cr 
&\times& \left\{k^2-m^2_f-2M_0f(m_f)\left[\frac{n\Lambda^2}{ 
n\Lambda^2-k^2}\right]^{2n}-M^2_0f^2\left[\frac{n\Lambda^2}{ 
n\Lambda^2-k^2}\right]^{4n}\right\}^{-1} 
\cr 
&\times&\left\{(k-P)^2-m^2_g-2M_0f(m_g)\left[\frac{n\Lambda^2}{ 
n\Lambda^2-(k-P)^2}\right]^{2n}-M^2_0f^2\left[\frac{n\Lambda^2}{ 
n\Lambda^2-(k-P)^2}\right]^{4n}\right\}^{-1}\nn\\&=&4\sqrt{ 
\eta_f\eta_g}\frac{(\beta k_--k^2_T)(\alpha k_--\gamma_f)^{3n}( 
\beta k_--\gamma_g)^{3n}}{\mathcal{D}_f\mathcal{D}_g}\cr 
&-&4\sqrt{\eta_f\eta_g}(\alpha  
k_-+\gamma_f)^n(\beta k_--\gamma_g)^n\frac{[m_f(\alpha 
  k_--\gamma_f)^{2n}+\eta_f][m_g(\beta 
  k_--\gamma_g)^{2n}+\eta_g]}{\mathcal{D}_f\mathcal{D}_g}, 
\label{aps2}   
\end{eqnarray} 
where we use the following abbreviations: 
\begin{eqnarray} 
&&\alpha=P_+,\,\,\,\,\beta=(u-1)P_+,\cr 
&&\gamma_f=k^2_T+n\Lambda^2,\,\,\,\,\gamma_g=(u-1)m^2_{\cal 
  M}+k^2_T+n\Lambda^2,\nn\\&&\delta_f=k^2_T+m^2_f,\,\,\,\, 
\delta_g=(u-1)m^2_{\cal M}+k^2_T+m^2_g,\cr 
&&\eta_f=M_0f(m_f)(n\Lambda^2)^{2n},\,\,\,\, 
\eta_f=M_0f(m_g)(n\Lambda^2)^{2n}.    
\end{eqnarray} 
$\mathcal{D}_{f,g}$ in the denominator of Eq.~(\ref{aps2}) reads: 
\begin{eqnarray}  
\mathcal{D}_{f,g}&=&\left[(\alpha k_--\delta_{f,g})(\alpha 
  k_--\gamma_{f,g})^{4n}-2m_{f,g}\eta_{f,g}(\alpha k_-- 
\gamma_{f,g})^{2n}-\eta^2_{f,g}+i\epsilon\right].  
\end{eqnarray} 
Finally, we have the following expression for the $\phi^p_{\cal 
 M}(u)$: 
\begin{eqnarray} 
\phi^p_{\cal 
  M}(u)&=&-\frac{4iN_c\sqrt{\eta_f\eta_g}(m_f+m_g)P_+}{m^2_{\cal 
    M}F^2_{\cal 
    M}}\int\frac{dk_-dk^2_T}{(2\pi)^3}\Bigg[\frac{(\beta  
  k_--k^2_T)(\alpha k_--\gamma_f)^{3n}(\beta 
  k_--\gamma_g)^{3n}}{\mathcal{D}_f\mathcal{D}_g}\cr 
&& \hspace{-1.6cm} -(\alpha  
k_-+\gamma_f)^n(\beta k_--\gamma_g)^n\frac{[m_f(\alpha 
  k_--\gamma_f)^{2n}+\eta_f][m_g(\beta 
  k_--\gamma_g)^{2n}+\eta_g]}{\mathcal{D}_f\mathcal{D}_g}\Bigg].   
\end{eqnarray} 
%------------------------------------------------- 
\subsection{Twist-3 pseudotensor type pseudoscalar meson DA} 
%------------------------------------------------- 
One can write the pseudotensor type twist-3 DA, using Eq.~(\ref{DAT}) as 
follows:   
\begin{eqnarray} 
\phi^{\sigma}_{\cal M}(u)&=&-\frac{6\sqrt{2}(m_f+m_g)}{m^2_{\cal M} 
F_{\cal M}}\int^{\infty}_{-\infty}\frac{d\tau}{\pi}\cr 
&& \times \int^u_0dv\,e^{-2iv\tau 
  P\cdot\hat{n}}\langle0|\bar{\psi}_f(\tau\hat{n})i(\rlap{/}{P} 
\rlap{/}{\hat{n}}-P\cdot\hat{n})\gamma_5\psi_g(-\tau\hat{n})| 
{\cal M}(P)\rangle.\nn\\ 
\label{apt2}   
\end{eqnarray} 
The trace in the matrix element of Eq.~(\ref{apt2}) can be derived as 
in the case of the pseudoscalar-type one.  Having performed the trace, 
we finally arrive at 
\begin{equation} 
\phi^{\sigma}_{\cal M}(u)=\frac{12iN_c(m_f+m_g)P^2_+}{m^2_{\cal M}F_{\cal  
M}F_{\pi}}\int_0^udv\int\frac{dk_-d^2k_T}{(2\pi)^2}\frac{k_-(\alpha 
  k_--\gamma_f)^{3n}(\beta k_--\gamma_g)^{3n}}{\mathcal{D}_f\mathcal{D}_g}.  
\label{apt3}   
\end{equation} 
%------------------------------------------------- 
 
\end{document}